# Exploiting Spatial Degrees of Freedom for High Data Rate Ultrasound Communication with Implantable Devices

Max L. Wang* and Amin Arbabian

*Department of Electrical Engineering, Stanford University, Stanford, California, 94305, USA*

We propose and demonstrate an ultrasonic communication link using spatial degrees of freedom to increase data rates for deeply implantable medical devices. Low attenuation and millimeter wavelengths make ultrasound an ideal communication medium for miniaturized low-power implants. While small spectral bandwidth has drastically limited achievable data rates in conventional ultrasonic implants, large spatial bandwidth can be exploited by using multiple transducers in a multiple-input/multiple-output system to provide spatial multiplexing gain without additional power, larger bandwidth, or complicated packaging. We experimentally verify the communication link in mineral oil with a transmitter and receiver 5 cm apart, each housing two custom-designed mm-sized piezoelectric transducers operating at the same frequency. Two streams of data modulated with quadrature phase-shift keying at 125 kbps are simultaneously transmitted and received on both channels, effectively doubling the data rate to 250 kbps with a measured bit error rate below $10^{-4}$. We also evaluate the performance and robustness of the channel separation network by testing the communication link after introducing position offsets. These results demonstrate the potential of spatial multiplexing to enable more complex implant applications requiring higher data rates.

Advances in miniaturized, wireless, and deeply implantable medical devices (IMDs) can enable coordinated closed-loop diagnostics and treatments for applications like neuromodulation and drug delivery.[1] As their capabilities become more complex and numerous, robust and low-power communication becomes increasingly important. Research into wireless networks in the body have largely focused on radio frequency (RF) communications.[2] However, a major challenge for the propagation of electromagnetic (EM) waves in the body is power absorption in tissue. The absorbed power is dissipated as heat leading to both health concerns and significant path loss.[3]

More recently, there has been increased interest in using ultrasonic waves for intra-body communication and power transfer.[2,4] At low MHz frequencies, reduced scattering due to relative homogeneity in tissue density and compressibility as well as low attenuation of about 1 dB·cm$^{-1}$·MHz$^{-1}$ in soft tissue allow ultrasonic waves to safely propagate much farther in tissue than EM waves.[1] The orders of magnitude slower propagation velocity of acoustic waves in the body (~1500 m/s) also results in millimeter wavelengths around a MHz, allowing for simpler circuits, beamforming capabilities, and smaller transducers.[1] On the other hand, the lower operating frequency and fundamentally smaller bandwidth drastically limit the achievable data rate and available modulation schemes. Required data rates can vary considerably depending on the application. For example, glucose monitoring may need kbps speeds while imaging/video may require Mbps speeds.[2] Recent studies have proposed and demonstrated different protocols with varying data rates for intra-body ultrasound communication.[5–8] However, there is a large disparity in achieved rates between systems using miniaturized IMDs (~10 kbps) and prototypes using large commercial wideband transducers (~Mbps). Major unavoidable challenges arise from miniaturization including degraded bandwidth, lower directivity, and limited processing capability stemming from packaging and power constraints.[4,9] These challenges have prevented ultrasonic IMDs from reaching the speeds necessary for more complex functionality.

Here, we propose the first ultrasonic intra-body communication system using spatial degrees of freedom in an acoustic multiple-input/multiple-output (MIMO) system to obtain multiplexing gain in order to increase data rates of conventional IMDs. We experimentally demonstrate the communication system with multiple mm-sized transducers so that our findings can be directly applied to miniaturized implants with minimal packaging constraints (Fig. 1). Both conventional and spatially multiplexed communications are demonstrated in 5 cm of mineral oil using a pseudo-random binary sequence (PRBS) at 125 kbps and 250 kbps respectively with a sufficiently low bit error rate (BER). Robustness of the system to location offsets is also evaluated theoretically and verified experimentally.

Spatial multiplexing is used in both fiber and free-space communications today to increase spectral efficiency by utilizing MIMO techniques.[10,11] Orthogonal modal bases along with multimode and multicore fibers are employed for spatial multiplexing in optical systems.[11–13] Because the waves are physically separated and/or spatially orthogonal, each stream can be exploited as an independent channel for communication.

*Corresponding author: maxlwang@stanford.edu

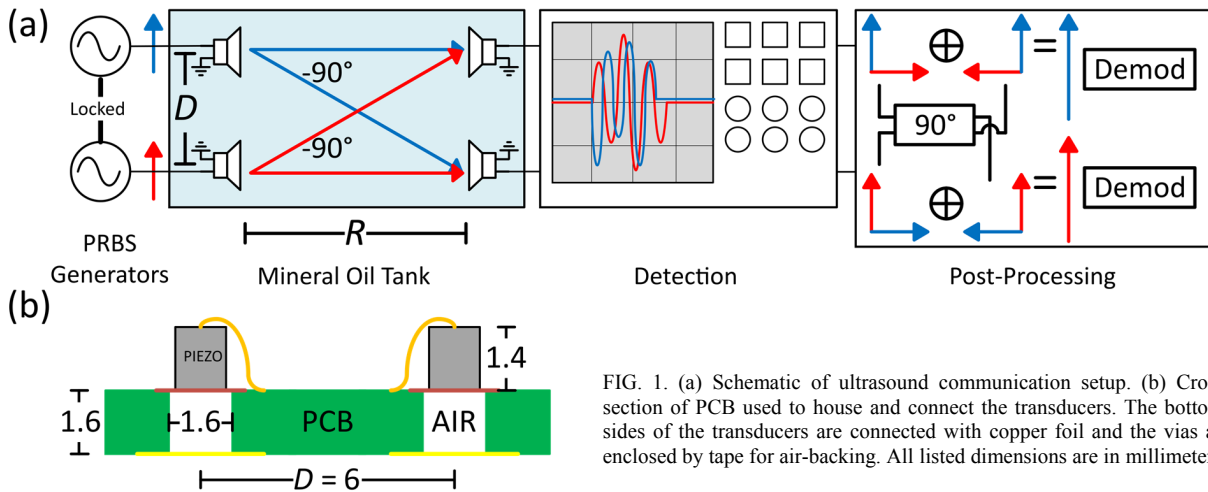

FIG. 1. (a) Schematic of ultrasound communication setup. (b) Cross-section of PCB used to house and connect the transducers. The bottom-sides of the transducers are connected with copper foil and the vias are enclosed by tape for air-backing. All listed dimensions are in millimeters.

RF communications have generally applied MIMO systems in scattering-rich environments with multiple paths from transmitters to receivers to sufficiently suppress spatial correlation between streams. With the appropriate channel separation network (CSN), the received signals can be decoupled, resulting in a theoretical channel capacity that scales with $N$ for an $N{\times}N$ system for the same power and bandwidth as a single-input/single-output (SISO) system.[10]

Orthogonality can also be obtained in a strongly line-of-sight (LOS) environment with minimal scattering, assuming sufficient transducer spacing.[10] This can intuitively be understood by using diffraction-limited optics. The ideal spacing needed to null the spatial interference for a transducer array is the same as the Rayleigh criterion for resolution, and is approximately

$$D \approx \sqrt{\frac{R \times \lambda}{N}} \qquad (1)$$

for $R \gg D$, where $R$ is the distance between transmitters and receivers, $\lambda$ is the carrier wavelength, and $N$ is the number of transmitters/receivers.[10] Previous demonstrations, all using EM waves, like those shown by Sheldon *et al.*[14] in LOS environments have only recently become feasible because carrier frequencies have been pushed high enough to allow reasonable antenna spacing for indoor and outdoor links. Fortunately, the short ultrasound wavelength significantly increases its spatial frequency capacity even in the MHz range, allowing spatial orthogonality to be attained in both LOS and non-LOS environments at centimeter distances. Therefore, the proposed ultrasonic links obtain their capacity from spatial rather than spectral bandwidth in the MHz range. This is important for intra-body communications because transmission distances can range from under a centimeter to multiple centimeters, and transmission paths may be strongly LOS due to low scattering and attenuation.

We implement a 2×2 ultrasonic communication link using miniaturized piezoelectric transducers to demonstrate spatial multiplexing gain as shown in Fig. 1(a). We designed the transducers with an aspect ratio near unity to reduce overall IMD volume. At these dimensions, the transducers can be modeled using length expander bar mode operation.[9] Bulk lead zirconate titanate 4 (PZT4) is diced to $1.08 \times 1.08 \times 1.44$ mm$^3$ dimensions to resonate near 1 MHz and provide ~2 k$\Omega$ radiation resistance at resonance, achieving a good tradeoff between transmit and receive efficiency.[9] Fig. 1(b) shows the cross-section of the identical printed circuit boards (PCBs) designed for the receiver and transmitter. Two PZT4 transducers are bonded on each board 6 mm apart ($D = 6$ mm), corresponding to an $R$ of 5 cm at 1 MHz from Eq. 1. The transducers are positioned over large, sealed vias in the PCB to provide air-backing and increase their sensitivity.[9,15]

The boards are placed in a tank filled with mineral oil, which is used to mimic the acoustic impedance of the body, albeit with lower attenuation, while minimizing electrical parasitics. A linear stage is used to move the receiver relative to the transmitter. Two phase-locked arbitrary waveform generators produce two independent PRBS streams and modulate the carrier with quadrature phase-shift keying (QPSK) at 125 kbps per stream for a total data rate of 250 kbps. These streams are then used to drive the transmitters with a total output power of about 600 μW, resulting in average intensities well below the diagnostic ultrasound limit in the body ($I_{SPTA} = 7.2$ mW/mm$^2$).[16] The emitted pressure fields from the two transducers are simulated at 1 MHz in a medium with 1 dB/cm attenuation using Field II (Fig. 2).[17] Interference from the two sources produce grating lobes with a half-power beamwidth of about 7° or 6 mm at a distance of 5 cm, matching the predicted Rayleigh resolution from Eq. 1 independent of attenuation. The two receive transducers denoted by the red crosses at the top of the figure can be seen to simultaneously receive data from both transmit transducers. For the experiment, the voltages from the two receive transducers are recorded with an oscilloscope and the CSN is implemented in MATLAB by filtering, shifting, and combining the signals to minimize interference before demodulation (Fig. 1(a)).

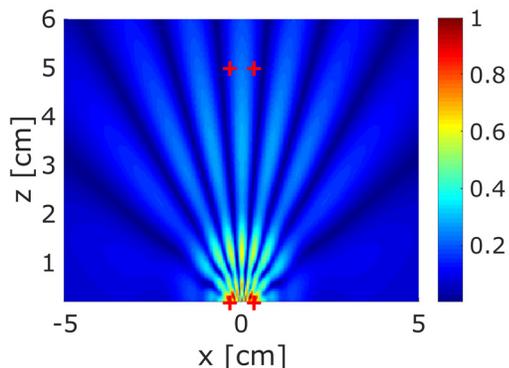

FIG. 2. Simulated normalized pressure fields emitted from two transducers operating in-phase at 1 MHz in a medium with 1 dB/cm attenuation. The red crosses denote the locations of the two transmitters (bottom) and two receivers (top).

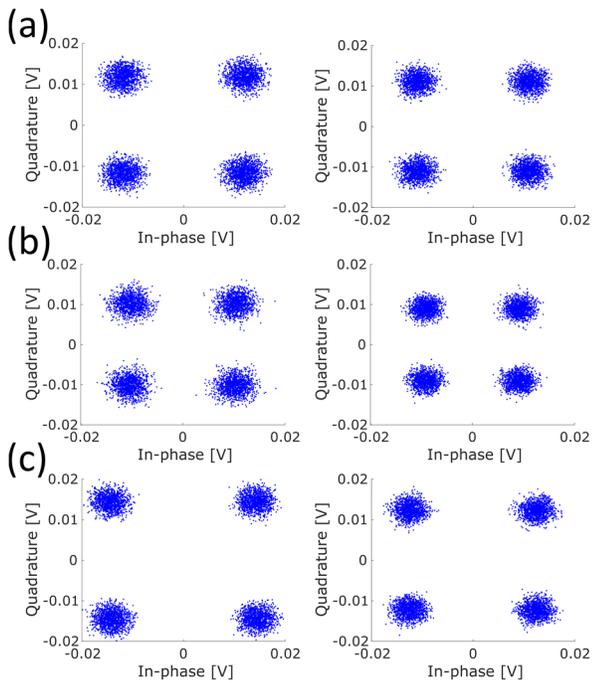

FIG. 3. Measured constellation diagrams of the received QPSK data from channel 1 (left) and channel 2 (right) for different operating modes. (a) Single transmitter and receiver (SISO) operating at 125 kbps. (b) Both channels operating simultaneously and decoupled with the CSN (MIMO) operating at 250 kbps. (c) Single transmitter and both receivers (SIMO) operating at 125 kbps for higher SNR.

TABLE I. Measured error vectors of each transducer configuration

| Configuration | Channel 1 [dB] | Channel 2 [dB] |
|---|---|---|
| SISO | -16.0 | -16.2 |
| MIMO | -14.8 | -15.5 |
|    Uncorrected offset | -12.5 | -13.2 |
|    Corrected offset | -14.8 | -15.1 |
| SIMO | -18.1 | -17.5 |

The ultrasound communication link is first tested with one transmit and one receive transducer at 125 kbps to characterize the performance of a conventional SISO communication link without interference. After verifying the operation of each transducer, two streams are transmitted simultaneously with the same total output power as the SISO configuration to verify the operation of the CSN and spatial multiplexing gain at 250 kbps. Communication between one transmitter and two receivers (or SIMO operation) is also tested at 125 kbps to demonstrate array gain through coherent combining of the received signals to increase the signal-to-noise ratio (SNR) in situations with less available power, more external interference, or increased attenuation. Fig. 3 shows the measured constellation diagrams from demodulating the received voltages in each of these operating modes, demonstrating the successful channel separation in the MIMO case and higher SNR in the SIMO case. The error vector magnitude (EVM), a common metric of received signal quality using the ratio of the measured deviation from the reference point to the power of the reference, is calculated from these constellation diagrams and is summarized in Table I.[18] The SISO and MIMO configurations show similar EVM while the SIMO configuration reduces the EVM consistent with higher SNR and absence of interference. The BER is tested to be $< 10^{-4}$ for each channel in all operation modes based on repeated acquisition of 500-bit PRBS packets.

We evaluate the robustness of our simple CSN by introducing offsets into the system. Ideally, the streams can be completely decoupled and the signal-to-interference ratio (SIR) is infinite. However, this is only true at a single frequency and distance. For any real system, noise and position offsets are non-zero, and there will be residual interference across streams because the channels are no longer exactly orthogonal, decreasing the SIR and SNR. The offsets are commonly corrected for in spatially-multiplexed communication systems with a reconfigurable CSN but a small power penalty is incurred from the resultant imperfect signal combining.[19] Fig. 4(a) shows the SIR as a function of phase offset (equivalent to location or frequency offset) as well as the penalty incurred with a reconfigured CSN to cancel the interference. To experimentally verify the robustness of spatial multiplexing to a position offset, $R$ is increased to 6.0 cm, corresponding to a phase offset of 15°. Without changing the CSN, the EVM of the received signals increases as can be seen by the blue circles in Fig. 4(b) and in Table I. However, by accounting for this position offset, the SIR can again be improved by increasing the phase shift in the CSN with negligible power penalty as can be seen in red in the same figure.

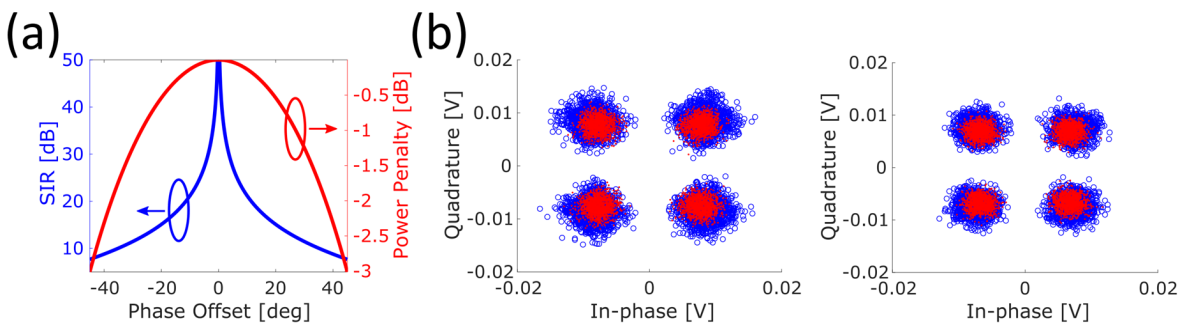

FIG. 4. (a) Theoretical robustness of the CSN to phase offset. For a static CSN (blue, left), the SIR decreases as offset increases because interference is not completely suppressed. If the CSN is reconfigured to suppress the interference (red, right), the signal power decreases because the channels are no longer completely orthogonal. (b) Constellation diagrams of the received data from channel 1 (left) and channel 2 (right) with a 1.0 cm offset with a static CSN (blue circles) and reconfigured CSN (red) showing improved EVM.

In conclusion, we have demonstrated that spatial degrees of freedom can be exploited with ultrasonic transducers to multiply data rates for intra-body communication. We experimentally verified these techniques with a 2×2 system using QPSK modulation for a combined data rate of 250 kbps, but this could easily be changed to a different modulation scheme as spatial multiplexing can complement many other communication techniques. Data rates can be further improved with a variety of methods including larger arrays, higher carrier frequencies, and error correction, as well as more sophisticated pre- and post-processing. For example, a 9×9 ultrasonic link using QPSK modulation at 4 MHz would only require an area of about 7×7 mm$^2$ and be able to support a 5 Mbps communication link. By using a mm-dimension transmitter and receiver, these findings can be directly applicable to a miniaturized implantable system, enabling wireless networks of implants and higher data rates for many potential applications.

The authors acknowledge the support from the Defense Advanced Research Projects Agency Young Faculty Award (Dr. Douglas Weber, program manager), the National Science Foundation CAREER award ECCS-1454107, the National Science Foundation Graduate Research Fellowship Program under Grant No. DGE-114747, and the 2015 Dr. Robert Noyce Stanford Graduate Fellowship. Special thanks to Mahmoud Sawaby and Ting-Chia Chang for valuable comments and suggestions.